\documentclass[12pt]{book}

\usepackage[dvips]{graphicx,color}
\usepackage{makeidx,tsukuba}

\makeauthorindex
\makeindex

\begin{document}

\BookTitle{\itshape The 28th International Cosmic Ray Conference}
\CopyRight{\copyright 2003 by Universal Academy Press, Inc.}
\pagenumbering{arabic}

\chapter{
Signal Fluctuations in the Auger Surface Detector
}

\author{%
%
%
Tokonatsu Yamamoto$^1$ for the Pierre Auger Collaboration$^2$\\
(1) Center for Cosmological Physics, University of Chicago\\
(2) Observatorio Pierre Auger, Malargue, 5613 Mendoze, Argentina
}

\section*{Abstract}

We measured the \v Cerenkov signal fluctuations in the water tanks of
the Pierre Auger Observatory (PAO). Two stations located near the center
of the 32-tank Engineering 
Array (EA) separated by 11 m were used for the purpose. At this
separation the stations sample nearly the same region of the air shower.
Sources of the signal fluctuations are discussed.

\section{Introduction}

The PAO is currently under construction in 
Mendoza Province, Argentina. Its main objective is the study of 
cosmic rays above 10 EeV. The Surface Detector (SD) will consist
of 1600 water tanks (10 $m^2\times$ 1.2 m deep) spaced
1.5 km apart to register the \v Cerenkov signal of a
sample of the Extensive Air Shower (EAS) particles. The fluctuations in
the signal recorded by the detectors directly affect the reconstruction
of the physical parameters.

The contributions to the total signal at ground level of the electromagnetic 
and muon components of the shower 
depend on the distance of the shower maximum to ground level
and the lateral distance of the tank from the core of the air shower.
The response of water \v Cerenkov tanks is different for these two
components. 
Therefore, the Poisson fluctuations will depend not only on the 
total signal, but also on primary energy, zenith angle, and distance of the shower core to the tank.
We call this ``sampling'' fluctuations.

The fluctuations we study depend not only on the physical processes in the
EAS but also on properties of the detector including stability and
calibration. Therefore an analysis of the signal characteristics is
basic to understanding the detector response and the data analysis.

We investigated the signal fluctuations based on the data obtained by
two closely located stations in the EA of the
PAO. Details of this 
analysis and the sources of signal fluctuations are discussed in this paper.

\section{Experiment}

The EA of the SD consists of 32 instrumented tanks separated by 1.5 km on a
triangular grid. It was successfully operated throughout most of
2002. Near the center of this array two stations are located at 11 m
separation. The stations are named Carmen and Miranda. This pair of
stations samples essentially the same region of the shower and therefore
is a useful tool for the study of the signal fluctuations
performed in this work. In addition to this pair there are two tanks on a
800 m spacing forming an equilateral triangle with the pair enabling
triggers at lower energies than with the standard 1500 m spacing.

To detect the \v Cerenkov light produced in the water each station has 3
PMTs located on the water surface. Calibration of each PMT is done with
single muon signals[1].

\begin{figure}[t]
  \begin{center}
    \includegraphics[height=14pc]{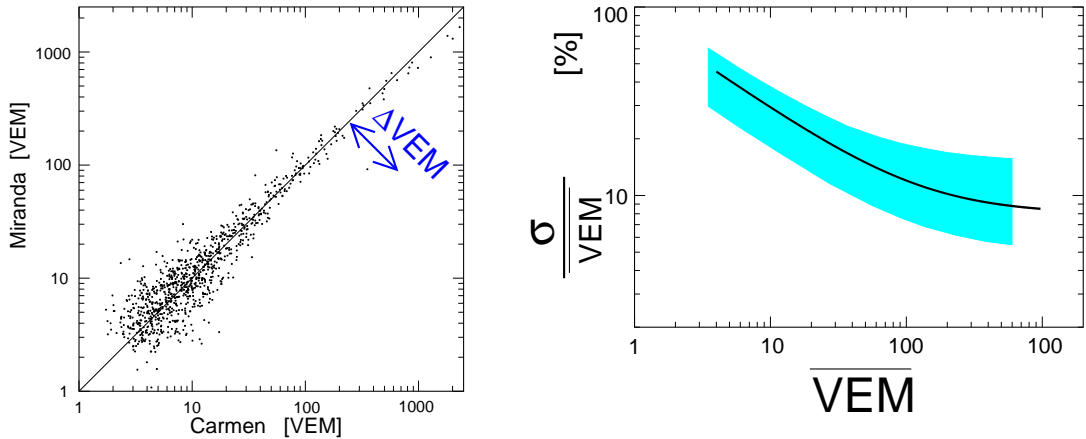}
  \end{center}
  \vspace{-0.5pc}
  \caption{Left: Correlation of the signals recorded in Carmen and
Miranda. The distance of each point to the solid line is the
fluctuation of each particular event.
Right: 
Signal fluctuation between Carmen and Miranda as a function of 
the  average of the two signals. 
The shaded area shows the uncertainty due to event selection. The solid line is 
representative of the data on average and indicates a fit using Equation
 (\ref{fm}) resulting in $A=0.08$ and $B=0.80$.
}
\label{fig:fig1}
\end{figure}

\begin{figure}[t]
  \begin{center}
    \includegraphics[height=13.5pc]{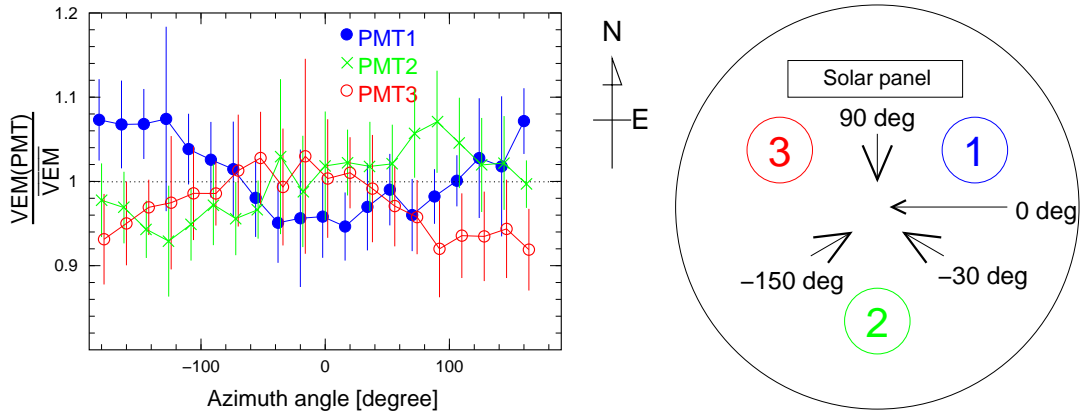}
  \end{center}
  \vspace{-0.5pc}
  \caption{Ratio of the signal of each PMT to the average of 3 PMTs
 as a function of azimuth angle of the air shower. Primary cosmic ray
 zenith angle smaller than 60 degrees and 10 or more $\overline{VEM}$
 are required. In this plot, a clear azimuthal effect is evident.
}
\label{fig:fig2}
\end{figure}

\section{Analysis}

We have analyzed all of the events recorded from May through October,
2002. About 1000 events with 4 or more stations triggered, including
Carmen and Miranda, survived after this event selection,
about half of which were triggered only by the 800 m triangle.
If the requirement is set for 5 or more stations triggered, the number of events is reduced to less than 500.

Left panel of Fig.1 shows the clear correlation of the signals 
recorded in Carmen and Miranda. The units are VEM, Vertical Equivalent
Muons, where 1 VEM is the average signal produced by a muon vertically
traversing the 
center of the tank. The distance of each point to the 
solid line is the fluctuation of each particular event ($\Delta VEM[i]$). 
The RMS of $\Delta VEM$ corresponds to the
signal fluctuation in one tank.

The sources of the signal fluctuations include:
(1) photo-statistics;
(2) calibration including detector stability and electrical noise;
(3) azimuthal effect;
(4) sampling fluctuations in the number of particles that hit each tank.

The Auger tanks have three measurements of the signal given by the 3 PMTs.
Particles from an air shower plunge into the tank and emit
\v Cerenkov light. The \v Cerenkov light is then diffusively reflected
by the walls of  
the tank liner and detected by the PMTs located at the top of the tank
(Fig.2).
Then the spread of the signal in the 3 PMTs will be driven by
Poisson fluctuations in the number of photoelectrons arriving to
the PMTs. These fluctuations are called ``Photo-statistic''.


If the tank is not a perfect diffuser, the number of \v Cerenkov photons
arriving at each PMT will be different depending on the incident angle
of the particles. \v Cerenkov photons 
falling directly onto the PMT without reflection from the tank walls
also contribute to this effect (Fig.2), which
we call the ``azimuthal effect''. 

Another source of fluctuation is what we called 
the ``lateral distribution effect''. The lateral distribution of particles per square meter is very steep
close to the core, and therefore the density changes very rapidly even
in distances as short as 11 m. Because of the uncertainty of the estimated
core location, this effect represents an additional uncertainty in our measurements of the fluctuations
near the core. This is one of the difficulties of this preliminary study.

Right panel of Fig.1 shows the fluctuations in 
the signal between Carmen and Miranda as a function of the 
average of the signal ($\overline{VEM}$).
We have analyzed the data based on different event selection
criteria.  The shaded area represents the uncertainty due to using these
different event selection criteria.
The fluctuations of smaller signals are dominated by the ``sampling''
fluctuations 
whereas for higher signals the fluctuations approach a constant 
fraction. It should be noted that the trigger efficiency will suppress
the fluctuations in lower $\overline{VEM}$ since the threshold of
the local trigger is adjusted to 3 VEM and we require both Carmen and
Miranda to trigger in this analysis. 
The anode output with a low amplification is used for large signals
(close to the shower axis) whereas the out from the last dynode is
strongly amplified (factor 30) to detect weaker signals.
Accuracy of the dynode to anode ratio,
which is about 5 \%, may dominate the constant fraction. 
Future on-line calibration system will reduce this fluctuation.

The functional form we propose to fit the signal fluctuations as
a function of VEM 
is the following:
\begin{equation}
\sigma_{\Delta VEM} = \sqrt{ ( VEM \times A )^2
 + ( \sqrt{VEM} \times B )^2}
\label{fm}
\end{equation}
where A is a constant parameter which is related to calibration accuracy
or detector stability and may be dominated by the accuracy of the 
Dynode to Anode ratio. $B$ 
is the parameter associated with the ``sampling'' fluctuations.  
Actually it should be a function of zenith angle and distance to the
shower core.
Our preliminary result, where we do not bin in angle or core distance, gives $A=0.08$ and $B=0.80$.

\section{Conclusions}

We have analyzed the data recorded by the Carmen and Miranda pair.
We have shown that the fluctuations of the 
mean signal in the tank can be explained with the functional form given
by Equation (\ref{fm}). 

The current statistics of the data are
not large enough to fit the parameters of Equation (\ref{fm}) as a function of 
zenith angle and distance to the shower core.
To investigate more detail, additional Carmen and Miranda pairs
and a larger data set are necessary. Both of them will be available in the
next stage of the Pierre Auger Surface Array.


\vspace{\baselineskip}

\re
1.\ Bertou X. for the Pierre Auger Collaboration, these proceedings



\endofpaper
\end{document}